\documentclass[fleqn,showpacs,showkeys]{revtex4}
\usepackage{graphicx}
\usepackage{amssymb}
\usepackage{amsmath}
\usepackage{bm}
\usepackage[flushleft]{caption2}

\begin{document}
\setcounter{page}{1}
\title{Octonionic electromagnetic and gravitational interactions and dark matter}
\author{Zihua Weng}
\email{xmuwzh@xmu.edu.cn.}
\affiliation{School
of Physics and Mechanical \& Electrical Engineering,
\\Xiamen University, Xiamen 361005, China}

\begin{abstract}
Based on the Maxwellian quaternionic electromagnetic theory, the
electromagnetic interaction, the gravitational interaction and their
coupling influence with the dark matter field in the octonionic
space are discussed. The paper reveals the close relationships of
dark matter field with electromagnetic field and gravitational
field. In the dark matter field, the research discloses the movement
properties of dark matter in certain conditions and its influence on
ordinary matter movements. In the electromagnetic field, the
variation of field energy density has direct effect on the force of
electric charge and current. In the modified gravitational field,
the near coplanarity, near circularity and corevolving of planetary
orbits are deduced from the equations set. The change of centrifugal
force of celestial body leads to fluctuation of revolution speed,
when the field energy density, celestial body's angular momentum in
its sense of revolution or its mass spatial distribution varied. The
results explain that some observed abnormal phenomena about
celestial bodies are caused by either modified gravitational field
or dark matter.
\end{abstract}



\maketitle

\section{\label{sec:level1}INTRODUCTION}

Nowadays, there still exist some movement phenomena of celestial
body which can't be explained by current gravitational theory.
Therefore, some scientists doubt the universality of current
gravitational theory, and then bring forward the modified
gravitational theory and hypothesis of dark matter to explain the
phenomena of abnormal rotation velocity of galaxy and the associated
'missing mass' etc. The hypothesis of dark matter believes that,
there exist not only various gleamy ordinary matters, but also one
kind of mysterious matter which exerts the gravitational pull, but
neither emits nor absorbs light \cite{bosma}. This kind of matter is
known as dark matter, which owns mass and has effect on ordinary
matter movements \cite{rubin}. Most scientists believe in its
existence and multiformity in the universe.

A new insight on the problem of the dark matter can be given by the
concept of the octonion space. The understanding of time and space
experiences an evolutional period \cite{lewis}. In 17th century, I.
Newton thought the time and space in nature were separate. And the
time and space were both independent of matter. In 20th century, A.
Einstein believed that, the time and space in nature were integrated
and commonly known as the spacetime. The spacetime related to the
movement status and matter. The precursors include R. Descartes and
A. Einstein etc. held the idea that space and time were the
extension of matter. And there was no such spacetime without matter
and hence no empty spacetime \cite{carmody}.

In the paper, we believe that electromagnetic and gravitational
interactions are different, equal and relative. And they possess
different and four-dimensional spacetimes respectively. These two
types of different and four-dimensional spacetimes are equal and
independent, and can not be superposed directly. The viewpoint can
be summarized as 'SpaceTime Equality Postulation', that is: each
fundamental interaction possesses its own spacetime, and all these
spacetimes are equal.

According to previous research results and the 'SpaceTime Equality
Postulation', the electromagnetic and gravitational interactions can
be described by their four-dimensional spacetimes. Based on the
conception of space verticality etc., two types of four-dimensional
spacetimes can be united into an eight-dimensional spacetime. In
eight-dimensional spacetime, the electromagnetic interaction and
gravitational interaction can be equally described. So the
characteristics of electromagnetic and gravitational interactions
can be described by a single eight-dimensional spacetime uniformly.

The paper modifies the gravitational theory of ordinary matter and
dark matter, and draws some conclusions which are consistent with
the Maxwellian electromagnetic theory and the Newtonian
gravitational theory. A few prediction associated with movement
feature of dark matter can be deduced, and some new particles can be
used for the candidate of dark matter.

\section{\label{sec:level1}Electromagnetic field and quaternion space}

The electromagnetic theory can be described with the quaternion
algebra \cite{elduque}. In the treatise on electromagnetic theory,
the quaternion algebra was first used by J. C. Maxwell to describe
the equations set and various properties of the electromagnetic
field \cite{mallett}. The spacetime, which is associated with
electromagnetic interaction and possesses the physics content, is
adopted by the quaternion space.

In the quaternion space of electromagnetic interaction, the radius
vector $\mathbb{R} = (r_0 , r_1 , r_2 , r_3)$ , and the base
$\mathbb{E}$ = ($1$, $\emph{\textbf{i}}_1$, $\emph{\textbf{i}}_2$,
$\emph{\textbf{i}}_3$). Where, $r_0 = ct$, $c$ is the speed of light
beam, and $t$ denotes the time.

The quaternion differential operator $\lozenge$ and its conjugate
operator $\lozenge^*$ are defined as
\begin{eqnarray}
\lozenge = \partial_0 + \emph{\textbf{i}}_1
\partial_1 + \emph{\textbf{i}}_2 \partial_2 +
\emph{\textbf{i}}_3 \partial_3 ~,~~ \lozenge^* = \partial_0
-\emph{\textbf{i}}_1 \partial_1 - \emph{\textbf{i}}_2
\partial_2 - \emph{\textbf{i}}_3 \partial_3~.
\end{eqnarray}
where, $\partial_i = \partial/\partial r_i$, i = 0, 1, 2, 3 . The mark (*) represents
the quaternion conjugate.

The electromagnetic field potential $\mathbb{A} (a_0 , a_1 , a_2 ,
a_3)$ is defined as
\begin{eqnarray}
\mathbb{A} = \lozenge \circ \mathbb{X} = a_0 + \emph{\textbf{i}}_1
a_1 + \emph{\textbf{i}}_2 a_2 + \emph{\textbf{i}}_3 a_3
\end{eqnarray}
where, $\mathbb{X}$ is the physical quantity.

\subsection{\label{sec:level2} Electromagnetic field strength}

The electromagnetic field strength $\mathbb{B} (b_0 , b_1 , b_2 ,
b_3)$ is defined as
\begin{eqnarray}
\mathbb{B}  =&& \lozenge \circ \mathbb{A}
\nonumber\\
 =&& b_{00} + (\emph{\textbf{i}}_1 b_{01} + \emph{\textbf{i}}_2 b_{02} + \emph{\textbf{i}}_3 b_{03})
+ (\emph{\textbf{i}}_1 b_{32} + \emph{\textbf{i}}_2 b_{13} +
\emph{\textbf{i}}_3 b_{21})
\nonumber\\
 = &&(\partial_0 a_0 - \partial_1 a_1 - \partial_2 a_2 - \partial_3
a_3 )
\nonumber\\
&& + \left\{\emph{\textbf{i}}_1 ( \partial_0 a_1 + \partial_1 a_0 )
+ \emph{\textbf{i}}_2 ( \partial_0 a_2 + \partial_2 a_0 ) +
\emph{\textbf{i}}_3 ( \partial_0 a_3 + \partial_3 a_0 ) \right\}
\nonumber\\
&& + \left\{\emph{\textbf{i}}_1 ( \partial_2 a_3 - \partial_3 a_2 )
+ \emph{\textbf{i}}_2 ( \partial_3 a_1 - \partial_1 a_3 ) +
\emph{\textbf{i}}_3 ( \partial_1 a_2 - \partial_2 a_1 ) \right\}
\end{eqnarray}
where, the first term in the right of equal mark is gauge
definition, the second term is electric field intensity $( b_{01} ,
b_{02} , b_{03} )$ and the third term is flux density $( b_{32} ,
b_{13} , b_{21} )$. Taking $b_0 = b_{00} = 0$, then the gauge
condition of field potential in electromagnetic field can be
achieved.

\begin{table}[h]
\caption{\label{tab:table1}The quaternion multiplication table.}
\begin{ruledtabular}
\begin{tabular}{ccccc}
$ $ & $1$ & $\emph{\textbf{i}}_1$  & $\emph{\textbf{i}}_2$ &
$\emph{\textbf{i}}_3$  \\
\hline $1$ & $1$ & $\emph{\textbf{i}}_1$  & $\emph{\textbf{i}}_2$ &
$\emph{\textbf{i}}_3$  \\
$\emph{\textbf{i}}_1$ & $\emph{\textbf{i}}_1$ & $-1$ &
$\emph{\textbf{i}}_3$  & $-\emph{\textbf{i}}_2$ \\
$\emph{\textbf{i}}_2$ & $\emph{\textbf{i}}_2$ &
$-\emph{\textbf{i}}_3$ & $-1$ & $\emph{\textbf{i}}_1$ \\
$\emph{\textbf{i}}_3$ & $\emph{\textbf{i}}_3$ &
$\emph{\textbf{i}}_2$ & $-\emph{\textbf{i}}_1$ & $-1$
\end{tabular}
\end{ruledtabular}
\end{table}

\subsection{\label{sec:level2} Electromagnetic field source}

The electromagnetic field source $\mathbb{S} (s_0 , s_1 , s_2 ,
s_3)$ is defined as
\begin{eqnarray}
- \mu \mathbb{S} = && \lozenge^* \circ \mathbb{B}
\nonumber\\
= &&(\partial_0 b_0 + \partial_1 b_1 + \partial_2 b_2 + \partial_3
b_3 )
\nonumber\\
&& + \left\{\emph{\textbf{i}}_1 ( \partial_0 b_1 - \partial_1 b_0 )
+ \emph{\textbf{i}}_2 ( \partial_0 b_2 - \partial_2 b_0 ) +
\emph{\textbf{i}}_3 ( \partial_0 b_3 - \partial_3 b_0 ) \right\}
\nonumber\\
&& + \left\{\emph{\textbf{i}}_1 ( \partial_3 b_2 - \partial_2 b_3 )
+ \emph{\textbf{i}}_2 ( \partial_1 b_3 - \partial_3 b_1 ) +
\emph{\textbf{i}}_3 ( \partial_2 b_1 - \partial_1 b_2 ) \right\}
\nonumber\\
= && \lozenge^2 ( a_0 + \emph{\textbf{i}}_1 a_1 +
\emph{\textbf{i}}_2 a_2 + \emph{\textbf{i}}_3 a_3 )
\end{eqnarray}
where, $\lozenge^2 = \lozenge^* \circ \lozenge $ .  $\mu$ is a
coefficient. $b_0 = 0$ . Taking $\mathbb{S} = ( s_0 +
\emph{\textbf{i}}_1 s_1 + \emph{\textbf{i}}_2 s_2 +
\emph{\textbf{i}}_3 s_3 )$, then equations set of field potential of
the electromagnetic field can be attained
\begin{eqnarray}
\lozenge^2( a_0 + \emph{\textbf{i}}_1 a_1 + \emph{\textbf{i}}_2 a_2
+ \emph{\textbf{i}}_3 a_3 )= - \mu( s_0 + \emph{\textbf{i}}_1 s_1 +
\emph{\textbf{i}}_2 s_2 + \emph{\textbf{i}}_3 s_3 )
\end{eqnarray}

\subsection{\label{sec:level2} Variation rate of field source}

The variation rate $\mathbb{J} (j_0 , j_1 , j_2 , j_3)$ of
electromagnetic field source is defined as
\begin{eqnarray}
\mathbb{J} = && \lozenge^* \circ \mathbb{S}
\nonumber\\
= &&(\partial_0 s_0 + \partial_1 s_1 + \partial_2 s_2 + \partial_3
s_3 )
\nonumber\\
&& + \left\{\emph{\textbf{i}}_1 ( \partial_0 s_1 - \partial_1 s_0 )
+ \emph{\textbf{i}}_2 ( \partial_0 s_2 - \partial_2 s_0 ) +
\emph{\textbf{i}}_3 ( \partial_0 s_3 - \partial_3 s_0 ) \right\}
\nonumber\\
&& + \left\{\emph{\textbf{i}}_1 ( \partial_3 s_2 - \partial_2 s_3 )
+ \emph{\textbf{i}}_2 (  \partial_1 s_3 - \partial_3 s_1 ) +
\emph{\textbf{i}}_3 (\partial_2 s_1 - \partial_1 s_2 ) \right\}
\end{eqnarray}

\begin{figure}[b]
\includegraphics{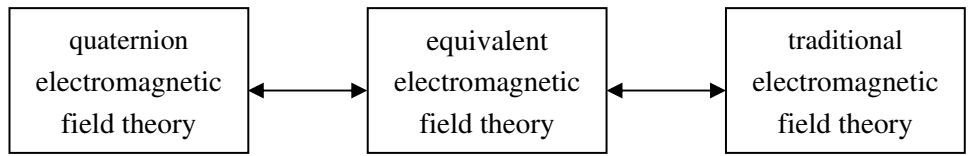}
\caption{\label{fig:epsart} Equivalent description in the
electromagnetic field theory. In the quaternion space, the various
properties among field potential, field strength and field source of
electromagnetic interaction can be described by quaternion. Those
conclusion and results are similar to that of equivalent
representation of the electromagnetic field.}
\end{figure}

Taking the gauge condition $j_0 = 0$, then the continuity equation
of field source of the electromagnetic field can be achieved.

In the quaternion space, various properties among the field
potential, field strength and field source of the electromagnetic
interaction can be described by the quaternion. Those conclusion and
results are similar to that of equivalent representation of the
electromagnetic field in Table 2. According to the 'SpaceTime
Equality Postulation', it can be deduced that the spacetime derived
from gravitational interaction is supposed to be the quaternion
space also. Given that the coupling influence of fundamental
interactions is neglected, the gravitational interaction is similar
to electromagnetic interaction and can also be described by
quaternion.

\begin{table*}[]
\begin{ruledtabular}
\caption{\label{tab:table1}Equivalent representation of the
electromagnetic field equation.}
\begin{tabular}{lll}
\textbf{} & traditional representation & equivalent representation\\
\hline
field potential & \emph{\textbf{A}}  & \emph{\textbf{A}}\\
                       &  $\varphi$         & $\varphi$ \\
\hline field strength & \emph{\textbf{B}} =
$\triangledown$$\times$\emph{\textbf{A}}
&  \emph{\textbf{B}} = $\triangledown$$\times$\emph{\textbf{A}} \\
& \emph{\textbf{E}} = $-$($\triangledown$$\varphi$ +
$\partial$\emph{\textbf{A}}/$\partial t$)
& \emph{\textbf{E}}$'$ = $\triangledown$$\varphi$ + $\partial$\emph{\textbf{A}}/$\partial t$\\
& \emph{\textbf{D}} = $\varepsilon$ \emph{\textbf{E}} &  \emph{\textbf{D}}$'$ = $\varepsilon$ \emph{\textbf{E}}$'$ \\
& \emph{\textbf{H}} = \emph{\textbf{B}}/$\mu$   & \emph{\textbf{H}} = \emph{\textbf{B}}/$\mu$  \\
\hline gauge condition & $\triangledown$$\cdot$\emph{\textbf{A}} +
$\partial$$(\varphi/c)$/$\partial$$(ct)$ = 0 &
$\triangledown$$\cdot$\emph{\textbf{A}} $-$
$\partial$$(\varphi/c)$/$\partial$$(ct)$ = 0 \\
\hline field source & $\triangledown^2\varphi -
\partial^2\varphi/\partial(ct)^2 = - \rho/\varepsilon$ &
$\triangledown^2\varphi +
\partial^2\varphi/\partial(ct)^2 = - \rho/\varepsilon$ \\
& $\triangledown^2\emph{\textbf{A}} -
\partial^2\emph{\textbf{A}}/\partial(ct)^2 = - \mu\emph{\textbf{j}}$ &
$\triangledown^2\emph{\textbf{A}} +
\partial^2\emph{\textbf{A}}/\partial(ct)^2 = - \mu\emph{\textbf{j}}$ \\
\hline Maxwell equation & $\triangledown\times\emph{\textbf{H}} =
\partial\emph{\textbf{D}}/\partial t + \emph{\textbf{j}}$
& $\triangledown\times\emph{\textbf{H}} =
\partial\emph{\textbf{D}}'/\partial t + \emph{\textbf{j}}$\\
& $\triangledown\times\emph{\textbf{E}} = -
\partial\emph{\textbf{B}}/\partial t $ & $\triangledown\times\emph{\textbf{E}}' =
\partial\emph{\textbf{B}}/\partial t $\\
& $\triangledown\cdot \emph{\textbf{D}}=\rho$
& $\triangledown\cdot \emph{\textbf{D}}$$'$=$ - \rho$\\
& $\triangledown\cdot \emph{\textbf{B}}=0$
& $\triangledown\cdot \emph{\textbf{B}}=0$ \\
\end{tabular}
\end{ruledtabular}
\label{tab:front}
\end{table*}

\section{\label{sec:level1}ELECTROMAGNETIC-GRAVITATIONAL FIELD AND OCTONION SPACE}

The electromagnetic and gravitational interactions are
interconnected, unified and equal. Both of them can be described in
the quaternion space. By means of the conception of space expansion
etc., two types of quaternion spaces can combine into an octonion
space \cite{bremner}. In the octonion space, various characteristics
of electromagnetic and gravitational interactions can be described
uniformly, and some correlative equations set can be obtained.

The base $\mathbb{E}_g$ = ($1$, $\emph{\textbf{i}}_1$,
$\emph{\textbf{i}}_2$, $\emph{\textbf{i}}_3$) of quaternion space of
gravitational interaction (G space, for short) is independent of the
base $\mathbb{E}_e$ of quaternion space of electromagnetic
interaction (E space, for short). Selecting $\mathbb{E}_e$ = ($1$,
$\emph{\textbf{i}}_1$, $\emph{\textbf{i}}_2$, $\emph{\textbf{i}}_3$)
$\circ$ $\emph{\textbf{I}}_0$ = ($\emph{\textbf{I}}_0$,
$\emph{\textbf{I}}_1$, $\emph{\textbf{I}}_2$,
$\emph{\textbf{I}}_3$). And then the bases $\mathbb{E}_g$ and
$\mathbb{E}_e$ constitute a single base $\mathbb{E}$ of the octonion
space.
\begin{equation}
\mathbb{E} = \mathbb{E}_g + \mathbb{E}_e = (1, \emph{\textbf{i}}_1,
\emph{\textbf{i}}_2, \emph{\textbf{i}}_3, \emph{\textbf{I}}_0,
\emph{\textbf{I}}_1, \emph{\textbf{I}}_2, \emph{\textbf{I}}_3)
\end{equation}

The radius vector $\mathbb{R} (r_0 , r_1 , r_2 , r_3 , R_0 , R_1 ,
R_2 , R_3 )$ in the octonion space is
\begin{equation}
\mathbb{R} = (r_0 + \emph{\textbf{i}}_1 r_1 + \emph{\textbf{i}}_2
r_2 + \emph{\textbf{i}}_3 r_3)+(\emph{\textbf{I}}_0 R_0 +
\emph{\textbf{I}}_1 R_1 + \emph{\textbf{I}}_2 R_2 +
\emph{\textbf{I}}_3 R_3)
\end{equation}
where, $r_0 = v_0 t$, $R_0 = V_0 T$. $v_0 = V_0 = c$ is the speed of
light beam, $t$ and $T$ denote the time.

The octonion algebra is the alternative algebra, so the octonions
$\mathbb{Q}_1$ and $\mathbb{Q}_2$  satisfy
\begin{eqnarray}
\mathbb{Q}_1 \circ ( \mathbb{Q}_1 \circ \mathbb{Q}_2 ) =
(\mathbb{Q}_1 \circ \mathbb{Q}_1) \circ \mathbb{Q}_2~,~~
\mathbb{Q}_1 \circ ( \mathbb{Q}_2 \circ \mathbb{Q}_2 ) =
(\mathbb{Q}_1 \circ \mathbb{Q}_2) \circ \mathbb{Q}_2~. \nonumber
\end{eqnarray}

The octonion differential operator $\lozenge$ and its conjugate
operator $\lozenge^*$ are defined as
\begin{subequations}
\begin{eqnarray}
&&\lozenge = \lozenge_g + \lozenge_e~,~~ \lozenge^* = \lozenge^*_g +
\lozenge^*_e ~;\label{a}
\\
&&\lozenge_g = \partial_{g0} + \emph{\textbf{i}}_1
\partial_{g1} + \emph{\textbf{i}}_2 \partial_{g2} +
\emph{\textbf{i}}_3 \partial_{g3}~,~~ \lozenge^*_g =
\partial_{g0} -\emph{\textbf{i}}_1 \partial_{g1} -
\emph{\textbf{i}}_2 \partial_{g2} - \emph{\textbf{i}}_3
\partial_{g3}~;\label{b}
\\
&&\lozenge_e = \emph{\textbf{I}}_0 \partial_{e0} +
\emph{\textbf{I}}_1
\partial_{e1} + \emph{\textbf{I}}_2
\partial_{e2} +\emph{\textbf{I}}_3 \partial_{e3}~,~~\lozenge^*_e = -
\emph{\textbf{I}}_0 \partial_{e0} - \emph{\textbf{I}}_1
\partial_{e1} - \emph{\textbf{I}}_2
\partial_{e2} - \emph{\textbf{I}}_3 \partial_{e3}~. \label{c}
\end{eqnarray}
\end{subequations}
where, $\partial_{gi}$ = $\partial$/$\partial r_i$ ; $\partial_{ei}$
= $\partial$/$\partial R_i$ ; i = 0, 1, 2, 3 . The mark (*) represents
the octonion conjugate.

The field potential $\mathbb{A} (a_0 , a_1 , a_2 , a_3 , A_0 , A_1 ,
A_2 , A_3 )$ in the electromagnetic-gravitational field is defined
as
\begin{eqnarray}
\mathbb{A} = \lozenge \circ \mathbb{X} = (a_0 +
\emph{\textbf{i}}_1 a_1 + \emph{\textbf{i}}_2 a_2 +
\emph{\textbf{i}}_3 a_3) + k_a (\emph{\textbf{I}}_0 A_0 +
\emph{\textbf{I}}_1 A_1 + \emph{\textbf{I}}_2 A_2 +
\emph{\textbf{I}}_3 A_3)
\end{eqnarray}
where, $\mathbb{X}=\mathbb{X}_g + k_x \mathbb{X}_e$ ; $\mathbb{X}_g$
and $\mathbb{X}_e$ are the physical quantity in G space and E space
respectively; $\mathbb{A}=\mathbb{A}_g + k_a \mathbb{A}_e$ ;
$\mathbb{A}_g = (a_0 , a_1 , a_2 , a_3)$ and $\mathbb{A}_e = (A_0 ,
A_1 , A_2 , A_3)$ are the field potential in G space and E space
respectively; $k_a$ and $k_x$ are coefficients.

\subsection{\label{sec:level2} Field strength}

The field strength $\mathbb{B} (b_0 , b_1 , b_2 , b_3 , B_0 , B_1 ,
B_2 , B_3 )$ is defined as
\begin{eqnarray}
\mathbb{B} = && \lozenge \circ \mathbb{A}
\nonumber\\
= && ( b_0 + \emph{\textbf{i}}_1 b_1 + \emph{\textbf{i}}_2 b_2 +
\emph{\textbf{i}}_3 b_3 ) + k_b (\emph{\textbf{I}}_0 B_0 +
\emph{\textbf{I}}_1 B_1 + \emph{\textbf{I}}_2 B_2 +
\emph{\textbf{I}}_3 B_3 )
\nonumber\\
= && [(\partial_{g0} a_0 - \partial_{g1} a_1 -
\partial_{g2} a_2 - \partial_{g3} a_3 )
\nonumber\\
&& + \emph{\textbf{i}}_1 ( \partial_{g0} a_1 + \partial_{g1} a_0 ) +
\emph{\textbf{i}}_2 ( \partial_{g0} a_2 + \partial_{g2} a_0 ) +
\emph{\textbf{i}}_3 ( \partial_{g0} a_3 + \partial_{g3} a_0 )
\nonumber\\
&& + \emph{\textbf{i}}_1 ( \partial_{g2} a_3 - \partial_{g3} a_2 ) +
\emph{\textbf{i}}_2 ( \partial_{g3} a_1 - \partial_{g1} a_3 ) +
\emph{\textbf{i}}_3 ( \partial_{g1} a_2 - \partial_{g2} a_1 ) ]
\nonumber\\
&& + [\emph{\textbf{I}}_0 (\partial_{e0} a_0 + \partial_{e1} a_1 +
\partial_{e2} a_2 + \partial_{e3} a_3 )
\nonumber\\
&& + \emph{\textbf{I}}_1 ( \partial_{e1} a_0 - \partial_{e0} a_1 ) +
\emph{\textbf{I}}_2 ( \partial_{e2} a_0 - \partial_{e0} a_2 ) +
\emph{\textbf{I}}_3 ( \partial_{e3} a_0 - \partial_{e0} a_3 )
\nonumber\\
&& + \emph{\textbf{I}}_1 (\partial_{e3} a_2 -  \partial_{e2} a_3 ) +
\emph{\textbf{I}}_2 (\partial_{e1} a_3 -  \partial_{e3} a_1 ) +
\emph{\textbf{I}}_3 (\partial_{e2} a_1 -  \partial_{e1} a_2 ) ]
\nonumber\\
&& + k_a [\emph{\textbf{I}}_0 (\partial_{g0} A_0 - \partial_{g1} A_1
- \partial_{g2} A_2 - \partial_{g3} A_3 )
\nonumber\\
&& + \emph{\textbf{I}}_1 ( \partial_{g1} A_0 + \partial_{g0} A_1 ) +
\emph{\textbf{I}}_2 ( \partial_{g2} A_0 + \partial_{g0} A_2 ) +
\emph{\textbf{I}}_3 ( \partial_{g3} A_0 + \partial_{g0} A_3 )
\nonumber\\
&& + \emph{\textbf{I}}_1 (\partial_{g3} A_2 -  \partial_{g2} A_3 ) +
\emph{\textbf{I}}_2 (\partial_{g1} A_3 -  \partial_{g3} A_1 ) +
\emph{\textbf{I}}_3 (\partial_{g2} A_1 -  \partial_{g1} A_2 ) ]
\nonumber\\
&& + k_a [- (\partial_{e0} A_0 + \partial_{e1} A_1 + \partial_{e2}
A_2 + \partial_{e3} A_3 )
\nonumber\\
&& + \emph{\textbf{i}}_1 (  \partial_{e0} A_1 - \partial_{e1} A_0 )
+ \emph{\textbf{i}}_2 (\partial_{e0} A_2 -  \partial_{e2} A_0 ) +
\emph{\textbf{i}}_3 (  \partial_{e0} A_3 - \partial_{e3} A_0 )
\nonumber\\
&& + \emph{\textbf{i}}_1 (\partial_{e3} A_2 - \partial_{e2} A_3 ) +
\emph{\textbf{i}}_2 (\partial_{e1} A_3 - \partial_{e3} A_1 ) +
\emph{\textbf{i}}_3 (\partial_{e2} A_1 - \partial_{e1} A_2 ) ]
\end{eqnarray}
where, the first two terms are the field strength of the
gravitational interaction, the last two terms are the field strength
of the electromagnetic interaction. $\mathbb{B}=\mathbb{B}_g + k_b
\mathbb{B}_e$ ; $\mathbb{B}_g = (b_1 , b_2 , b_3)$ and $\mathbb{B}_e
= (B_1 , B_2 , B_3)$ are the field strength in G space and E space
respectively; $k_b$ is a coefficient. Selecting gauge equations $b_0
= 0$ and $B_0 = 0$ can simplify definition of field strength in the
octonion space.

\begin{table}[h]
\caption{\label{tab:table1}The octonion multiplication table.}
\begin{ruledtabular}
\begin{tabular}{ccccccccc}
$ $ & $1$ & $\emph{\textbf{i}}_1$  & $\emph{\textbf{i}}_2$ &
$\emph{\textbf{i}}_3$  & $\emph{\textbf{I}}_0$  &
$\emph{\textbf{I}}_1$
& $\emph{\textbf{I}}_2$  & $\emph{\textbf{I}}_3$  \\
\hline $1$ & $1$ & $\emph{\textbf{i}}_1$  & $\emph{\textbf{i}}_2$ &
$\emph{\textbf{i}}_3$  & $\emph{\textbf{I}}_0$  &
$\emph{\textbf{I}}_1$
& $\emph{\textbf{I}}_2$  & $\emph{\textbf{I}}_3$  \\
$\emph{\textbf{i}}_1$ & $\emph{\textbf{i}}_1$ & $-1$ &
$\emph{\textbf{i}}_3$  & $-\emph{\textbf{i}}_2$ &
$\emph{\textbf{I}}_1$
& $-\emph{\textbf{I}}_0$ & $-\emph{\textbf{I}}_3$ & $\emph{\textbf{I}}_2$  \\
$\emph{\textbf{i}}_2$ & $\emph{\textbf{i}}_2$ &
$-\emph{\textbf{i}}_3$ & $-1$ & $\emph{\textbf{i}}_1$  &
$\emph{\textbf{I}}_2$  & $\emph{\textbf{I}}_3$
& $-\emph{\textbf{I}}_0$ & $-\emph{\textbf{I}}_1$ \\
$\emph{\textbf{i}}_3$ & $\emph{\textbf{i}}_3$ &
$\emph{\textbf{i}}_2$ & $-\emph{\textbf{i}}_1$ & $-1$ &
$\emph{\textbf{I}}_3$  & $-\emph{\textbf{I}}_2$
& $\emph{\textbf{I}}_1$  & $-\emph{\textbf{I}}_0$ \\
\hline $\emph{\textbf{I}}_0$ & $\emph{\textbf{I}}_0$ &
$-\emph{\textbf{I}}_1$ & $-\emph{\textbf{I}}_2$ &
$-\emph{\textbf{I}}_3$ & $-1$ & $\emph{\textbf{i}}_1$
& $\emph{\textbf{i}}_2$  & $\emph{\textbf{i}}_3$  \\
$\emph{\textbf{I}}_1$ & $\emph{\textbf{I}}_1$ &
$\emph{\textbf{I}}_0$ & $-\emph{\textbf{I}}_3$ &
$\emph{\textbf{I}}_2$  & $-\emph{\textbf{i}}_1$
& $-1$ & $-\emph{\textbf{i}}_3$ & $\emph{\textbf{i}}_2$  \\
$\emph{\textbf{I}}_2$ & $\emph{\textbf{I}}_2$ &
$\emph{\textbf{I}}_3$ & $\emph{\textbf{I}}_0$  &
$-\emph{\textbf{I}}_1$ & $-\emph{\textbf{i}}_2$
& $\emph{\textbf{i}}_3$  & $-1$ & $-\emph{\textbf{i}}_1$ \\
$\emph{\textbf{I}}_3$ & $\emph{\textbf{I}}_3$ &
$-\emph{\textbf{I}}_2$ & $\emph{\textbf{I}}_1$  &
$\emph{\textbf{I}}_0$  & $-\emph{\textbf{i}}_3$
& $-\emph{\textbf{i}}_2$ & $\emph{\textbf{i}}_1$  & $-1$ \\
\end{tabular}
\end{ruledtabular}
\end{table}

\subsection{\label{sec:level2} Field source}

The field source $\mathbb{S}(s_0 , s_1 , s_2 , s_3 , S_0 , S_1 , S_2
, S_3 )$ is defined as
\begin{eqnarray}
- \mu \mathbb{S}= && \lozenge^* \circ \mathbb{B}
\nonumber\\
= && [(\partial_{g0} b_0 + \partial_{g1} b_1 +
\partial_{g2} b_2 + \partial_{g3} b_3 )
\nonumber\\
&& + \emph{\textbf{i}}_1 ( \partial_{g0} b_1 - \partial_{g1} b_0 ) +
\emph{\textbf{i}}_2 ( \partial_{g0} b_2 - \partial_{g2} b_0 ) +
\emph{\textbf{i}}_3 ( \partial_{g0} b_3 - \partial_{g3} b_0 )
\nonumber\\
&& + \emph{\textbf{i}}_1 (\partial_{g3} b_2 -  \partial_{g2} b_3 ) +
\emph{\textbf{i}}_2 ( \partial_{g1} b_3 - \partial_{g3} b_1 ) +
\emph{\textbf{i}}_3 (\partial_{g2} b_1 -  \partial_{g1} b_2 ) ]
\nonumber\\
&& + [- \emph{\textbf{I}}_0 (\partial_{e0} b_0 + \partial_{e1} b_1 +
\partial_{e2} b_2 + \partial_{e3} b_3 )
\nonumber\\
&& + \emph{\textbf{I}}_1 (\partial_{e0} b_1 -  \partial_{e1} b_0 ) +
\emph{\textbf{I}}_2 (\partial_{e0} b_2 -  \partial_{e2} b_0 ) +
\emph{\textbf{I}}_3 (\partial_{e0} b_3 -  \partial_{e3} b_0 )
\nonumber\\
&& + \emph{\textbf{I}}_1 (\partial_{e2} b_3 - \partial_{e3} b_2 ) +
\emph{\textbf{I}}_2 (\partial_{e3} b_1 - \partial_{e1} b_3 ) +
\emph{\textbf{I}}_3 (\partial_{e1} b_2 - \partial_{e2} b_1 ) ]
\nonumber\\
&& + k_b [\emph{\textbf{I}}_0 (\partial_{g0} B_0 + \partial_{g1} B_1
+ \partial_{g2} B_2 + \partial_{g3} B_3 )
\nonumber\\
&& + \emph{\textbf{I}}_1 (\partial_{g0} B_1 - \partial_{g1} B_0) +
\emph{\textbf{I}}_2 (\partial_{g0} B_2 - \partial_{g2} B_0) +
\emph{\textbf{I}}_3 (\partial_{g0} B_3 - \partial_{g3} B_0)
\nonumber\\
&& + \emph{\textbf{I}}_1 (\partial_{g2} B_3 - \partial_{g3} B_2) +
\emph{\textbf{I}}_2 (\partial_{g3} B_1 - \partial_{g1} B_3) +
\emph{\textbf{I}}_3 (\partial_{g1} B_2 - \partial_{g2} B_1) ]
\nonumber\\
&& + k_b [(\partial_{e0} B_0 + \partial_{e1} B_1 + \partial_{e2} B_2
+ \partial_{e3} B_3 )
\nonumber\\
&& + \emph{\textbf{i}}_1 (\partial_{e1} B_0 - \partial_{e0} B_1) +
\emph{\textbf{i}}_2 (\partial_{e2} B_0 - \partial_{e0} B_2) +
\emph{\textbf{i}}_3 (\partial_{e3} B_0 - \partial_{e0} B_3)
\nonumber\\
&& + \emph{\textbf{i}}_1 (\partial_{e2} B_3 - \partial_{e3} B_2) +
\emph{\textbf{i}}_2 (\partial_{e3} B_1 - \partial_{e1} B_3) +
\emph{\textbf{i}}_3 (\partial_{e1} B_2 - \partial_{e2} B_1) ]
\nonumber\\
= && - \mu_g^g (s_0^g + \emph{\textbf{i}}_1 s_1^g +
\emph{\textbf{i}}_2 s_2^g + \emph{\textbf{i}}_3 s_3^g)
- \mu_g^e(\emph{\textbf{I}}_0 s_0^e + \emph{\textbf{I}}_1 s_1^e
+\emph{\textbf{I}}_2 s_2^e +\emph{\textbf{I}}_3 s_3^e)
\nonumber\\
&& - k_b \mu_e^g ( S_0^g + \emph{\textbf{i}}_1 S_1^g +
\emph{\textbf{i}}_2 S_2^g + \emph{\textbf{i}}_3 S_3^g) - k_b \mu_e^e
(\emph{\textbf{I}}_0 S_0^e + \emph{\textbf{I}}_1 S_1^e
+\emph{\textbf{I}}_2 S_2^e + \emph{\textbf{I}}_3 S_3^e)
\end{eqnarray}
where, the first two terms are the definition of momentum, and the
last two terms are the definition of current; $\mu$, $\mu_e^e$,
$\mu_e^g$, $\mu_g^e$ and $\mu_g^g$ are coefficients.

The physical quantities in the quaternion space need to be extended
to that in the octonion space. In the octonion space, the definition
of physical quantity should be equal and consistent in the
electromagnetic and gravitational interactions.

The velocity $\mathbb{V}(v_0 , v_1 , v_2 , v_3 , V_0 , V_1 , V_2 ,
V_3 )$ of particle is defined as
\begin{eqnarray}
\mathbb{V} = \partial\mathbb{R}/\partial t = (v_0 +
\emph{\textbf{i}}_1 v_1 + \emph{\textbf{i}}_2 v_2 +
\emph{\textbf{i}}_3 v_3) + (\emph{\textbf{I}}_0 V_0 +
\emph{\textbf{I}}_1 V_1 + \emph{\textbf{I}}_2 V_2 +
\emph{\textbf{I}}_3 V_3)
\end{eqnarray}
where, the first term is the velocity in G space, and the second
term is the velocity in E space.

In the octonion space, the charge and mass of particles are defined
as follows
\begin{subequations}
\begin{eqnarray}
&& s_0^g + \emph{\textbf{i}}_1 s_1^g + \emph{\textbf{i}}_2 s_2^g
+\emph{\textbf{i}}_3 s_3^g = Q_g^g (v_0 + \emph{\textbf{i}}_1 v_1
+\emph{\textbf{i}}_2 v_2 + \emph{\textbf{i}}_3 v_3)\label{a}
\\
&& \emph{\textbf{I}}_0 s_0^e + \emph{\textbf{I}}_1 s_1^e
+\emph{\textbf{I}}_2 s_2^e + \emph{\textbf{I}}_3 s_3^e = Q_g^e
(\emph{\textbf{I}}_0 V_0 + \emph{\textbf{I}}_1 V_1 +
\emph{\textbf{I}}_2 V_2 + \emph{\textbf{I}}_3 V_3)\label{b}
\\
&& S_0^g + \emph{\textbf{i}}_1 S_1^g + \emph{\textbf{i}}_2 S_2^g
+\emph{\textbf{i}}_3 S_3^g = Q_e^g (v_0 + \emph{\textbf{i}}_1 v_1
+\emph{\textbf{i}}_2 v_2 + \emph{\textbf{i}}_3 v_3)\label{c}
\\
&& \emph{\textbf{I}}_0 S_0^e + \emph{\textbf{I}}_1 S_1^e
+\emph{\textbf{I}}_2 S_2^e + \emph{\textbf{I}}_3 S_3^e = Q_e^e
(\emph{\textbf{I}}_0 V_0 + \emph{\textbf{I}}_1 V_1
+\emph{\textbf{I}}_2 V_2 + \emph{\textbf{I}}_3 V_3)\label{d}
\end{eqnarray}
\end{subequations}
where, the first two equations are the definition
of mass $Q_g^g$ and $Q_g^e$ respectively; the last two equations are
the definition of charge $Q_e^g$ and $Q_e^e$ respectively. The
charges $Q_g^g$, $Q_g^e$, $Q_e^g$ and $Q_e^e$ can be known as the
'general charge' uniformly.

In the octonion space, there are four types of field sources in the
electromagnetic-gravitational field. There exist two parts of field
sources in the electromagnetic interaction, one part lies in G space
($1$, $\emph{\textbf{i}}_1$, $\emph{\textbf{i}}_2$,
$\emph{\textbf{i}}_3$) and the other in E space
($\emph{\textbf{I}}_0$, $\emph{\textbf{I}}_1$,
$\emph{\textbf{I}}_2$, $\emph{\textbf{I}}_3$). So does the
gravitational interaction.

\subsection{\label{sec:level2} Force and angular momentum}

In the electromagnetic-gravitational field, the force can be defined
by the linear momentum $\mathbb{P} = \mu\mathbb{S}/\mu^g_g$, field
strength $\mathbb{B}$, and operator $\lozenge$. The field force
$\mathbb{F}$ and variation rate $\mathbb{J}$ of field source are
defined as
\begin{equation}
\mathbb{F} = \mathbb{B}^* \circ \mathbb{P}~,~~ \mathbb{J} =
\lozenge^* \circ \mathbb{P}~.
\end{equation}

The definition of variation rate $\mathbb{J}$ of field source is
different from that of field force $\mathbb{F}$. But both of them
have effects on the matter particle. The field force $\mathbb{F}$
includes the Lorentz force \cite{caponio}, while the variation rate
$\mathbb{J}$ of field source includes the inertia force
\cite{lynden-bell}.

The angular momentum $\mathbb{M}$ is defined as ($k_{rx}$ is a
coefficient)
\begin{equation}
\mathbb{M} = (\mathbb{R}+k_{rx}\mathbb{X}) \circ \mathbb{P}
\end{equation}

The $\mathbb{M}$ is defined by the linear momentum $\mathbb{P}$,
field strength $\mathbb{B}$, radius vector $\mathbb{R}$, physical
quantity $\mathbb{X}$ and operator $\lozenge$, and includes the
angular momentum, work, kinetic energy, potential energy and force
moment etc.

The work and moment derived from the field force are defined
uniformly as
\begin{equation}
\mathbb{W}_1 = \mathbb{B} \circ (\mathbb{R} \circ \mathbb{P})
+\mathbb{B} \circ (k_{rx}\mathbb{X} \circ \mathbb{P})
\end{equation}
where, the physical quantity $\mathbb{B} \circ (\mathbb{X} \circ
\mathbb{P}) $ is a new part of the energy.

The kinetic energy, potential energy and moment derived from the
variation rate of field source are defined as
\begin{equation}
\mathbb{W}_2 = \lozenge \circ (\mathbb{R} \circ \mathbb{P})
+\lozenge \circ (k_{rx}\mathbb{X} \circ \mathbb{P})
\end{equation}
where, the physical quantity $\lozenge \circ (\mathbb{X} \circ
\mathbb{P})$ is the part of potential energy.

Both definitions of energy $\mathbb{W}_1$ and $\mathbb{W}_2$ should
exclude the potential energy part $\mathbb{A} \circ \mathbb{P}$, if
the definition of angular momentum is confined in $\mathbb{M} =
 \mathbb{R} \circ \mathbb{P}$ . Hence this definition of angular
momentum is inexact apparently. As shown in Aharonov-Bohm experiment
\cite{aharonov}, the field potential $\mathbb{A} = \lozenge \circ
\mathbb{X}$ has effects on the energy $\lozenge \circ (\mathbb{X}
\circ \mathbb{P})$, which includes the potential energy $\mathbb{A}
\circ \mathbb{P}$ . And the change of field potential can impact the
movement of field source.

\section{\label{sec:level1}EQUATIONS SET OF ELECTROMAGNETIC-GRAVITATIONAL FIELD}

In the octonion electromagnetic-gravitational field, two types of
the forces can be written together as follows from Eq.(15) ($\alpha
= c$ is a coefficient)
\begin{equation}
\mathbb{Z} = \mathbb{F} + \alpha \mathbb{J}= (\mathbb{B} + \alpha
\lozenge)^* \circ \mathbb{P}
\end{equation}

And two types of energies can also be written as follows from
Eqs.(17) and (18)
\begin{equation}
\mathbb{W} = \mathbb{W}_1 + \alpha \mathbb{W}_2 = (\mathbb{B} +
\alpha \lozenge) \circ \mathbb{M}
\end{equation}

The unified definitions of force Eq.(19) and energy Eq.(20) show
that, previous equations set of the quaternion electromagnetic field
and octonion electromagnetic-gravitational field should be
supplemented and generalized. It also shows that the octonion
differential operator $\lozenge$ needs to be generalized to a new
operator $(\lozenge+\mathbb{B}/\alpha)$. The physical
characteristics of electromagnetic-gravitational field can be
described uniformly from many aspects.

\subsection{\label{sec:level2} Field source equation}

The extended definition of force Eq.(19) shows that, the field
source $\mathbb{S}$ needs to be revised and generalized to the form
of following equation from Eq.(12)
\begin{equation}
\mu \mathbb{S} = - (\mathbb{B}/ \alpha + \lozenge)^* \circ \mathbb{B}
= (\mu\mathbb{S})' - \mathbb{B}^* \circ \mathbb{B}/\alpha
\end{equation}
where, $(\mu\mathbb{S})'= - \lozenge^*\circ\mathbb{B}$ is the
definition of the field source in Eq.(12) in the Maxwellian
electromagnetic theory or the Newtonian gravitational theory.

As one part of field source $\mathbb{S}$, the term $(\mathbb{B}^*
\circ \mathbb{B}/\alpha)$ is directly proportional of field energy
density $(\mathbb{B}^* \circ \mathbb{B}/\mu_g^g)$. The force-balance
equation can be obtained when the force $\mathbb{Z}$ = 0 from
Eq.(19). And the variation of term $(\mathbb{B}^* \circ
\mathbb{B}/\alpha)$ will be one kind of necessary portion of force
$\mathbb{Z}$ .

\begin{table*}[h]
\begin{ruledtabular}
\caption{\label{tab:table1}The subfield types of
electromagnetic-gravitational field.}
\begin{tabular}{lll}
\textbf{}Operator      & Gravitational Interaction                & Electromagnetic Interaction\\
\hline
operator $\lozenge_g$  & gravitational-gravitational subfield     & electromagnetic-gravitational subfield\\
of G space             & G mass, $Q_g^g$                          & G charge, $Q_e^g$ \\
                       & intermediate particle, $\gamma_g^g$      & intermediate particle, $\gamma_e^g$\\
\hline
operator $\lozenge_e$  & gravitational-electromagnetic subfield   & electromagnetic-electromagnetic subfield\\
of E space             & E mass, $Q_g^e$                          & E charge, $Q_e^e$ \\
                       & intermediate particle, $\gamma_g^e$      & intermediate particle, $\gamma_e^e$\\
\end{tabular}
\end{ruledtabular}
\label{tab:front}
\end{table*}

\subsection{\label{sec:level2} Power equation}

The extended definition of energy Eq.(20) shows that, the power
$\mathbb{N}$ needs to be revised and generalized to the form of the
following equation.
\begin{equation}
\mathbb{N} = (\mathbb{B} + \alpha \lozenge)^* \circ \mathbb{W}
\end{equation}

In the electromagnetic-gravitational field, the physical quantity
$\mathbb{X}$ has effect on field potential $\lozenge \circ
\mathbb{X}$ , angular momentum $ \mathbb{X} \circ \mathbb{P}$ ,
energy $\mathbb{B} \circ (\mathbb{X} \circ \mathbb{P})$ and power
$(\mathbb{B} \circ \mathbb{B}^* ) \circ (\mathbb{X} \circ
\mathbb{P})$ etc. The introduction of physical quantity $\mathbb{X}$
makes the definition of angular momentum $\mathbb{M}$ and energy
$\mathbb{W}$ more integrated, and the theory more self-consistent.

In Eq.(20), the conservation of angular momentum in
electromagnetic-gravitational field can be gained when $\mathbb{W} =
0$. And the energy conservation of electromagnetic-gravitational
field can be attained from Eq.(22) when $\mathbb{N} = 0$.

\subsection{\label{sec:level2} Dark matter}

In the octonion spacetime, there exist four types of subfields and
their field source of electromagnetic-gravitational field. In the
Table 4, the electromagnetic-gravitational (E-G) subfield and
gravitational-gravitational (G-G) subfield are 'electromagnetic
field' and 'modified gravitational filed' respectively. Their
general charges are G charge and G mass respectively. The
electromagnetic-electromagnetic (E-E) subfield and
gravitational-electromagnetic (G-E) subfield are both long range
fields and candidates of the 'dark matter field'. Their general
charges (E charge and E mass) are candidates of 'dark matter'. The
physical features of the dark matter meet the requirement of
Eqs.(10), (11), (16), (19), (20), (21) and (22) .

In the general charge ($Q_g^g$, $Q_g^e$, $Q_e^g$ and $Q_e^e$) and
intermediate particle ($\gamma_g^g$, $\gamma_g^e$, $\gamma_e^g$ and
$\gamma_e^e$), two types of general charges (familiar charge $Q_e^g$
and mass $Q_g^g$) and one type of intermediate particle (photon
$\gamma_e^g$) have been found. Hence the remaining two types of
general charges and three types of intermediate particles are left
to be found in Table 4. From Eq.(12), we find that the
electromagnetic-gravitational field can transform one form of
intermediate particle into another. In other words, the photon
$\gamma_e^g$ can be transformed into the $\gamma_g^g$, $\gamma_g^e$,
or $\gamma_e^e$ .

The particles of ordinary matter (electron and proton etc.) possess
the G charge together with G mass. The particles of dark matter may
possess the E charge with E mass, or G mass with E charge, etc. It
can be predicted that field strength of
electromagnetic-electromagnetic and gravitational-electromagnetic
subfields must be very weak and a little less than that of the
gravitational-gravitational subfield. Otherwise they should been
detected for a long time. Hence the field strength of
electromagnetic-electromagnetic and gravitational-electromagnetic
subfields may be equal, and a little less than that of
gravitational-gravitational subfield.

\begin{table*}[h]
\caption{\label{tab:table3} The comparison between ordinary matter
with dark matter.}
\begin{ruledtabular}
\begin{tabular}{ccccc}
&\multicolumn{2}{c}{Ordinary~Matter}&\multicolumn{2}{c}{Dark~Matter}\\
 \hline
 subfield & G-G~subfield & E-G~subfield & G-E~subfield & E-E~subfield \\
 field~potential & $\lozenge_g \circ \mathbb{X}_g$ & $\lozenge_g \circ \mathbb{X}_e$
 & $\lozenge_e \circ \mathbb{X}_g$ & $\lozenge_e \circ \mathbb{X}_e$\\
 field~strength & $\lozenge_g \circ \mathbb{A}_g$ & $\lozenge_g \circ \mathbb{A}_e$
 & $\lozenge_e \circ \mathbb{A}_g$ & $\lozenge_e \circ \mathbb{A}_e$\\
 field~source & $\lozenge_g^* \circ \mathbb{B}_g$ & $\lozenge_g^* \circ \mathbb{B}_e$
 & $\lozenge_e^* \circ \mathbb{B}_g$ & $\lozenge_e^* \circ \mathbb{B}_e$\\
\end{tabular}
\end{ruledtabular}
\end{table*}

\section{\label{sec:level1}EXPERIMENTS AND PHENOMENA}

The field sources (current and momentum) in Eq.(19) stays in the
force-balance status in the electromagnetic-gravitational field when
$\mathbb{Z} = 0$. The influence between the 'dark matter field'
(electromagnetic-electromagnetic and gravitational-electromagnetic
subfields) and the 'modified gravitational field'
(gravitational-gravitational subfield) on the force $\mathbb{Z}$ can
be discussed with this feature. The conservation of angular momentum
in Eq.(20) in the electromagnetic-gravitational field can be
achieved when $\mathbb{W} = 0$. The influence between the 'modified
gravitational field' and 'dark matter field' on the angular momentum
and energy can be considered by this property. It is simpler and
more convenient to represent the field source $(\mu\mathbb{S})'$ as
$\mu\mathbb{S}$ in the following discussion.

\begin{figure}[t]
\includegraphics{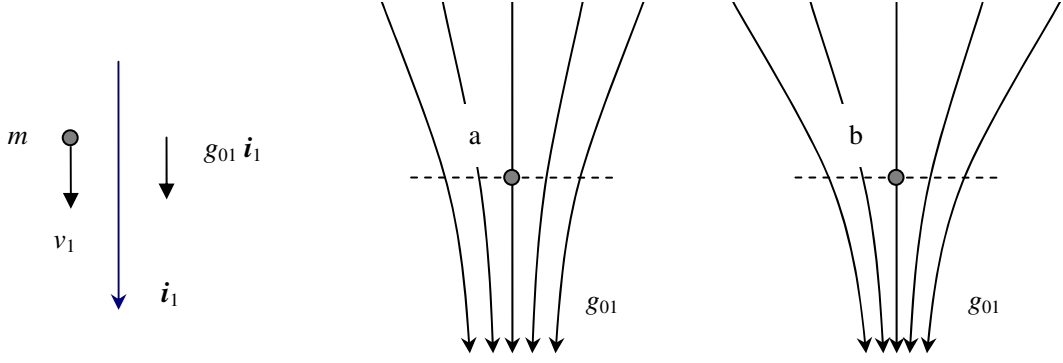}
\caption{\label{fig:epsart} Linear movement. The change of spatial
distribution of mass \emph{m} will affect free-fall movement of
particle \emph{m}. The linear movement status of particle \emph{m}
in point \emph{a} and \emph{b} are different, though the field
strength are equivalent in those points.}
\end{figure}

\subsection{\label{sec:level2} Linear movement}

In the gravitational-gravitational subfield ($g_{01} , 0 , 0$) in
Fig.2, the particle \emph{m} is in free-fall movement and its
momentum is ($s_0^g , s_1^g , 0 , 0$). When the particle \emph{m} is
in weightlessness status,
\begin{eqnarray}
\mathbb{Z} = (\emph{\textbf{i}}_1 g_{01} + \alpha \lozenge)^* \circ
\left\{\mu_g^g (s_0^g + \emph{\textbf{i}}_1 s_1^g) - g_{01}^2
/\alpha \right\}/\mu^g_g = 0
\end{eqnarray}
then eight equations can be obtained from Eq.(23). And its equation
in $\emph{\textbf{i}}_1$ direction is
\begin{eqnarray}
\mu_g^g g_{01} s_0^g + \alpha \mu_g^g (\partial_{g1} s_0^g -
\partial_{g0} s_1^g ) - (\partial_{g1} g_{01}^2 + g_{01}^3 /\alpha
)= 0
\end{eqnarray}

When the last term of the above equation is zero, we can achieve
\begin{eqnarray}
g_{01} s_0^g = c (\partial_{g0} s_1^g - \partial_{g1} s_0^g)
\end{eqnarray}
where, the physical quantity can be written in familiar form,
$g_{01} = g/c$ , $s_0^g = mc$ , $s_1^g = mv_1$.

The above shows that, when the spatial distribution of the mass
\emph{m} changes, as one part of inertia force, $\partial_{g1}
s_0^g$ will affect free-fall movement of particle \emph{m}. The
omitted term ($\partial_{g1} g_{01}^2 + g_{01}^3/\alpha$) in Eq.(24)
will influence free-fall movement of particle \emph{m} also. Then
the linear movement status of particle \emph{m} in the point
\emph{a} and \emph{b} are different in Fig.2, though the field
strength $g_{01}$ in the point \emph{a} and \emph{b} are equal.

\begin{figure}[t]
\includegraphics{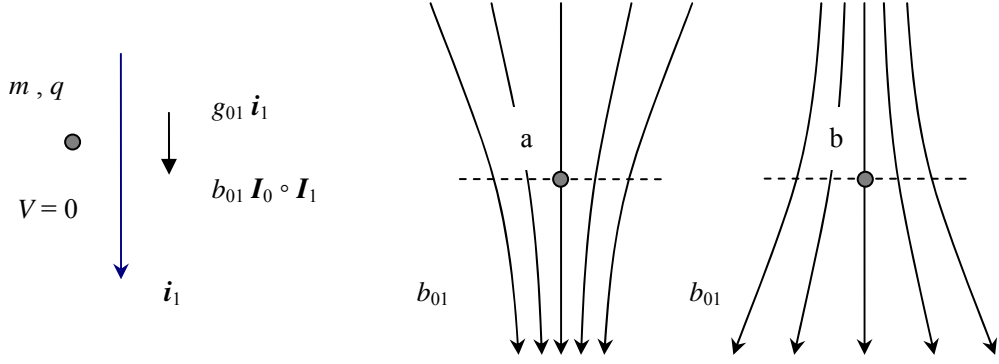}
\caption{\label{fig:epsart} Stationary status. The transform of
energy density in the electromagnetic-gravitational field will
affect the force-balance of charged particle including the
stationary status of electric charged particle. In consequence the
stationary status of electric charged particle \emph{q} in point
\emph{a} and \emph{b} will be different.}
\end{figure}

\subsection{\label{sec:level2} Stationary status}

In the area, where coexist the electromagnetic-gravitational
subfield ($b_{01} , 0 , 0$) and gravitational-gravitational subfield
($g_{01} , 0 , 0$). The electric charged particle \emph{q} is in the
stationary status in the area in Fig.3, its momentum is ($s_0^g , 0
, 0 , 0$) and its current is ($S_0^g , 0 , 0 , 0$). When the
particle \emph{q} stays in the force-balance status
\begin{eqnarray}
\mathbb{Z} = \left\{(\emph{\textbf{i}}_1 g_{01} + k_a
\emph{\textbf{I}}_1 b_{01}) + \alpha \lozenge \right\}^*
 \circ \left\{(\mu_g^g s_0^g + k_b \mu_e^g \emph{\textbf{I}}_0
S_0^g) - (g_{01}^2 + k_a^2 b_{01}^2)/\alpha \right\}/\mu^g_g = 0
\end{eqnarray}
so its equation in $\emph{\textbf{i}}_1$ direction is
\begin{eqnarray}
(\mu_g^g g_{01} s_0^g - k_b \mu_e^g k_a b_{01} S_0^g ) + \alpha (
\mu_g^g \partial_{g1} s_0^g - k_b \mu_e^g \partial_{e1} S_0^g) -
\partial_{g1} (g_{01}^2 + k_a^2 b_{01}^2) - g_{01} (g_{01}^2 + k_a^2
b_{01}^2) /\alpha = 0
\end{eqnarray}

The last two terms of the above equation show that, the transform of
energy density in the electromagnetic-gravitational and the
gravitational-gravitational subfields will affect the force-balance
of charged particle. When the sum of above last three terms equals
to zero, and $k_a k_b =  \mu_g^g / \mu_e^g$ , we can conclude from
above
\begin{eqnarray}
g_{01} s_0^g - b_{01} S_0^g = 0 \nonumber
\end{eqnarray}
where, the physical quantity can be written in the familiar form,
$b_{01} = E/c, g_{01} = g/c, s_0^g = mc, S_0^g = qc$ .

The last three terms in Eq.(27) will impact the stationary status of
electric charged particle. So the stationary status of electric
charged particle \emph{q} in point \emph{a} and \emph{b} will be
different in Fig.3, though the field strength $b_{01}$ in point
\emph{a} and \emph{b} are equal.

\subsection{\label{sec:level2} Planar circumference movement of dark matter}

In the electromagnetic-gravitational subfield (0, 0, $b_{12}$), the
charged particle \emph{q} is in the planar circumference movement in
Fig.4. Its momentum is ($s_0^g , s_1^g , s_2^g , 0$), while its two
types of currents are ($S_0^e , S_1^e , S_2^e , 0$) and ($S_0^g ,
S_1^g , S_2^g , 0$) respectively. When the particle \emph{q} is in
the force-balance status,
\begin{eqnarray}
\mathbb{Z} = &&  (\emph{\textbf{I}}_3 k_a b_{12} + \alpha
\lozenge)^* \circ \nonumber
\\
&& \left\{\mu_g^g (s_0^g + \emph{\textbf{i}}_1 s_1^g +
\emph{\textbf{i}}_2 s_2^g) + k_b \mu_e^e( \emph{\textbf{I}}_0 S_0^e
+\emph{\textbf{I}}_1 S_1^e + \emph{\textbf{I}}_2 S_2^e) + k_b
\mu_e^g(S_0^g + \emph{\textbf{i}}_1 S_1^g + \emph{\textbf{i}}_2
S_2^g) - k_a^2 b_{12}^2 /\alpha \right\}/\mu^g_g = 0
\end{eqnarray}
so its equations in $\emph{\textbf{i}}_1$ and $\emph{\textbf{i}}_2$
directions are respectively
\begin{subequations}
\begin{eqnarray}
&& k_a k_b b_{12} \mu_e^e S_2^e + \alpha \mu_g^g (-\partial_{g0}
s_1^g + \partial_{g1} s_0^g - \partial_{g3}s_2^g ) - \partial_{g1}
(k_a b_{12})^2
\nonumber\\
&& + \alpha k_b \mu_e^e (\partial_{e0} S_1^e - \partial_{e1} S_0^e
+\partial_{e3} S_2^e ) + \alpha k_b \mu_e^g ( -\partial_{g0} S_1^g
-\partial_{g1} S_0^g + \partial_{g3}S_2^g)=0 \label{a}
\\
&& - k_a k_b b_{12} \mu_e^e S_1^e + \alpha \mu_g^g (-\partial_{g0}
s_2^g + \partial_{g2} s_0^g + \partial_{g3}s_1^g ) - \partial_{g2}
(k_a b_{12})^2
\nonumber\\
&& + \alpha k_b \mu_e^e ( \partial_{e0} S_2^e - \partial_{e2} S_0^e
-\partial_{e3} S_1^e ) + \alpha k_b \mu_e^g ( -\partial_{g0} S_2^g
+\partial_{g2} S_0^g + \partial_{g3}S_1^g)= 0 \label{b}
\end{eqnarray}
\end{subequations}

It makes out that field source ($S_0^e , S_1^e , S_2^e , 0$) of
'dark matter field' has direct influence on the movement of field
sources ($S_0^g , S_1^g , S_2^g , 0$) of familiar 'electromagnetic
field'.

When the sum of last three terms in above each equation equals to 0,
and $ k_a k_b = \mu_g^g / \mu_e^g $, the force-balance equations of
charged particle, which are in the circumference movement in
electromagnetic field and modified gravitational field, can be
achieved from above
\begin{subequations}
\begin{eqnarray}
&& (\mu^g_g \mu^e_e /\mu^g_e) b_{12} S_2^e + c (-\partial_{g0}s_1^g
+ \partial_{g1} s_0^g -\partial_{g3}s_2^g )=0 \label{a}\\
&& (\mu^g_g \mu^e_e /\mu^g_e)  b_{12} S_1^e - c (-\partial_{g0}
s_2^g + \partial_{g2} s_0^g +\partial_{g3} s_1^g )=0 \label{b}
\end{eqnarray}
\end{subequations}

The above equations show that, as the parts of inertia force, the
terms $(\partial_{g1} s_0^g - \partial_{g3}s_2^g )$ and
$(\partial_{g2} s_0^g + \partial_{g3} s_1^g) $ will affect the
planar circumference movement of charged particle. The terms
$\partial_{g1} b_{12}^2$ in Eq.(29a) and $\partial_{g2} b_{12}^2$ in
Eq.(29b) will affect the planar circumference movement of charged
particle \emph{q} also.

\begin{figure}[h]
\begin{minipage}[t]{0.5\linewidth}
\setcaptionwidth{3in}
\includegraphics[]{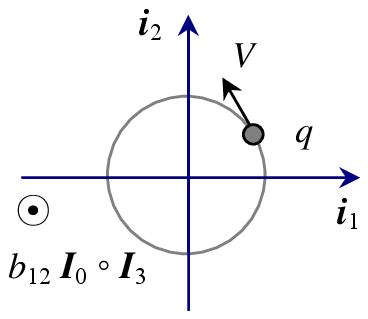}
\caption{Planar circumference movement. The field source of 'dark
matter field' has direct influence on the movement of field source
of familiar 'electromagnetic field'.} \label{fig:side:a}
\end{minipage}%
\begin{minipage}[t]{0.5\linewidth}
\setcaptionwidth{3.5in}
\includegraphics[]{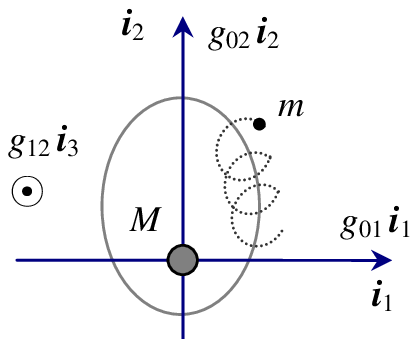}
\caption{Near circularity and rotation. The omitted 'magnetic' part
of gravitational-gravitational subfield strength affects the
movement of particle, and is one of important factors resulting in
rotation of celestial bodies.} \label{fig:side:b}
\end{minipage}
\end{figure}

\subsection{\label{sec:level2} Near circularity and rotation}

The large-mass object \emph{M} generates the field strength
($g_{01}, g_{02}, g_{12}$) of the gravitational-gravitational
subfield in its equator plane. The particle \emph{m} moves around
\emph{M} in planar helix, and its momentum in orbit plane is
($s_0^g, s_1^g, s_2^g, 0$). The orbit plane of particle \emph{m} is
superposed with the equator plane of large-mass object \emph{M} in
Fig.5. When particle \emph{m} stays in force-balance status,
\begin{eqnarray}
\mathbb{Z} =&& (\emph{\textbf{i}}_1 g_{01} + \emph{\textbf{i}}_2
g_{02} + \emph{\textbf{i}}_3 g_{12} + \alpha \lozenge)^* \circ
\left\{\mu_g^g (s_0^g + \emph{\textbf{i}}_1 s_1^g
+\emph{\textbf{i}}_2 s_2^g) - B^2 /\alpha \right\}/\mu_g^g = 0
\end{eqnarray}
so its equations in $\emph{\textbf{i}}_1$ and $\emph{\textbf{i}}_2$
directions are
\begin{subequations}
\begin{eqnarray}
&& \mu_g^g (g_{01}s_0^g + g_{12}s_2^g) + \alpha \mu_g^g
(-\partial_{g0} s_1^g + \partial_{g1} s_0^g - \partial_{g3}s_2^g) -
\partial_{g1}B^2 - g_{01}B^2/\alpha =0
\label{a}\\
&& \mu_g^g (g_{02}s_0^g - g_{12}s_1^g) + \alpha \mu_g^g
(-\partial_{g0} s_2^g + \partial_{g2} s_0^g +
\partial_{g3}s_1^g) - \partial_{g2}B^2 - g_{02}B^2/\alpha =0
\label{b}
\end{eqnarray}
\end{subequations}
where, $B^2 = g_{01}^2 + g_{02}^2 + g_{12}^2$ .

When the sum of last two terms in above each equation is equal
approximately to zero, we can attain
\begin{subequations}
\begin{eqnarray}
&& (g_{01}s_0^g + g_{12}s_2^g) + c (-\partial_{g0} s_1^g +
\partial_{g1} s_0^g - \partial_{g3}s_2^g) =0
\label{a} \\
&& (g_{02}s_0^g - g_{12}s_1^g) + c (-\partial_{g0} s_2^g
+\partial_{g2} s_0^g + \partial_{g3}s_1^g) =0 \label{b}
\end{eqnarray}
\end{subequations}

When $g_{12}$ = 0, we can learn that the particle \emph{m} moves
around object \emph{M} in planar ellipse from above equations. That
is the near circularity of planets. When $g_{01} = g_{02}$ = 0, it
can be found that the particle \emph{m} moves in planar
circumference from above equations. That is the rotation of planets.
Therefore, in gravitational-gravitational subfield ($g_{01}, g_{02},
g_{12}$) , the centroid movement of particle \emph{m} is the
superposition of these two types of movements. That is, the center
of planar circumference movement of particle moves in planar
ellipse. When the particle \emph{m} moves around big mass \emph{M},
the movement of particle \emph{m} is in revolution, rotation and
swing because of weakness and nonuniform distribution of $g_{12}$
\cite{ohtsuki}. It illustrates that, the omitted 'magnetic' part
$g_{12}$ of gravitational-gravitational subfield strength affects
the movement of planet \emph{m}, and is one of important factors
resulting in rotation of celestial bodies.

As shown above, the terms ($\partial_{g1} s_0^g  -  \partial_{g3}
s_2^g$) and ($\partial_{g2} s_0^g + \partial_{g3} s_1^g$), as
certain parts of inertia force, will affect the planar helix
movement of particle in Fig.6. In the solar system, when spatial
distribution of planetary mass and field energy density are
invariable, or when the planetary momentum is invariable along the
direction perpendicular to the Ecliptic Plane, the movement of
planet is consistent with the Newtonian gravitational theory.
Contrarily, in the Milky Way Galaxy, when the spatial distribution
of solar system mass and field energy density are variable, or when
the angular momentum of the solar system is variable along the
direction perpendicular to the Galactic Equator Plane, the
centrifugal force of solar system will change, and then the movement
of solar system will betray the familiar gravitational theory. If
the change of mass, energy density and momentum lead to counteract
the certain parts of centrifugal force, it will result in the
revolution speed increase to generate enough centrifugal force to
balance gravitational interaction of the Milky Way Galaxy. Therefore
the inferences may solve inconsistency between the overspeed
rotation of galaxy and deficiency of gravitation in some extent
\cite{pashkevich}.

\begin{figure}
\begin{minipage}[t]{0.5\linewidth}
\setcaptionwidth{3.3in}
\includegraphics[]{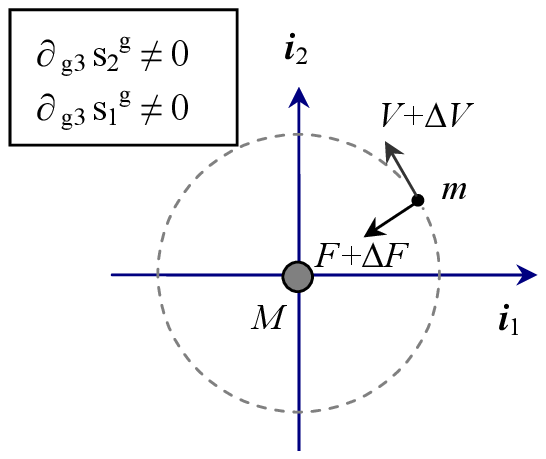}
\caption{Overspeed revolution. If the change of mass, energy density
and momentum lead to counteract the certain parts of centrifugal
force, it will result in the revolution speed increase to generate
enough centrifugal force to balance gravitational interaction.}
\label{fig:side:a}
\end{minipage}%
\begin{minipage}[t]{0.5\linewidth}
\setcaptionwidth{3.3in}
\includegraphics[]{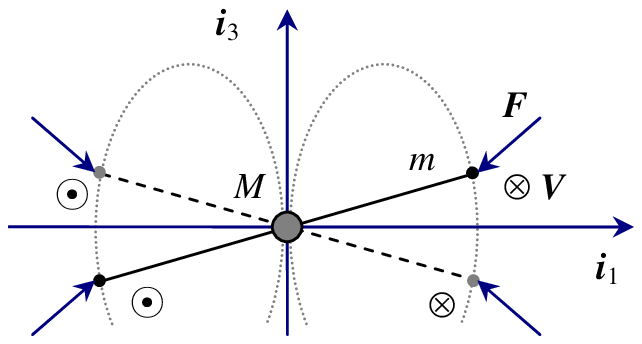}
\caption{Near coplanarity and corevolving. When the sense of
revolution of particle \emph{m} is in the same sense of rotation of
object \emph{M}, the movement of particle \emph{m} is attracted by
the force which points to the orbit plane, hence the movement of
particle \emph{m} is stable.} \label{fig:side:b}
\end{minipage}
\end{figure}

\subsection{\label{sec:level2} Near coplanarity and corevolving}

On the orbit plane of particle \emph{m} which momentum is ($s_0^g ,
s_1^g , s_2^g , s_3^g$), the gravitational-gravitational subfield
($g_{01}+g_{23} , g_{02}+g_{31} , g_{03}+g_{12}$) and the
electromagnetic-gravitational subfield ($b_{01}+b_{23} ,
b_{02}+b_{31} , b_{03}+b_{12}$) are generated by the large-mass
object \emph{M} which owns rotation and charge in Fig.7. When the
particle \emph{m}, which moves around \emph{M} in spatial helix, is
in the force-balance status,
\begin{eqnarray}
\mathbb{Z} = && [\emph{\textbf{i}}_1 (g_{01} + g_{23}) +
\emph{\textbf{i}}_2 (g_{02}+g_{31}) + \emph{\textbf{i}}_3 (g_{03}+
g_{12}) \nonumber\\
&&+ k_a \left\{\emph{\textbf{I}}_1 (b_{01} + b_{23}) +
\emph{\textbf{I}}_2 (b_{02} + b_{31}) + \emph{\textbf{I}}_3 (b_{03}
+ b_{12})\right\} + \alpha \lozenge ]^* \circ
\nonumber\\
&&\left\{\mu_g^g (s_0^g + \emph{\textbf{i}}_1 s_1^g +
\emph{\textbf{i}}_2 s_2^g + \emph{\textbf{i}}_3 s_3^g ) - B^2/\alpha
\right\}/\mu_g^g = 0
\end{eqnarray}
so its equations in $\emph{\textbf{i}}_1$ , $\emph{\textbf{i}}_2$
and $\emph{\textbf{i}}_3$ directions respectively are,
\begin{subequations}
\begin{eqnarray}
&& \mu_g^g \left\{(g_{01} + g_{23}) s_0^g + (g_{02} + g_{31}) s_3^g
- (g_{03} + g_{12})s_2^g \right\}
\nonumber \\
&& + \alpha \mu_g^g (-\partial_{g0} s_1^g +
\partial_{g1} s_0^g + \partial_{g2}s_3^g - \partial_{g3}s_2^g) -
\partial_{g1}B^2 - (g_{01}+g_{23})B^2/\alpha = 0
\label{a}\\
&& \mu_g^g \left\{(g_{02} + g_{31}) s_0^g - (g_{01} + g_{23}) s_3^g
+ (g_{03} + g_{12})s_1^g \right\}
\nonumber \\
&& + \alpha \mu_g^g (-\partial_{g0} s_2^g -
\partial_{g1} s_3^g + \partial_{g2}s_0^g + \partial_{g3}s_1^g) -
\partial_{g2}B^2 - (g_{02}+g_{31})B^2/\alpha = 0
\label{b}\\
&& \mu_g^g \left\{(g_{03} + g_{12}) s_0^g + (g_{01} + g_{23}) s_2^g
- (g_{02} + g_{31})s_1^g \right\}
\nonumber \\
&& + \alpha \mu_g^g (-\partial_{g0} s_3^g +
\partial_{g1} s_2^g - \partial_{g2}s_1^g + \partial_{g3}s_0^g) -
\partial_{g3}B^2 - (g_{03}+g_{12})B^2/\alpha = 0
\label{c}
\end{eqnarray}
\end{subequations}
where,
$B^2=(g_{01}+g_{23})^2+(g_{02}+g_{31})^2+(g_{03}+g_{12})^2
+k_a^2\left\{(b_{01}+b_{23})^2+(b_{02}+b_{31})^2+(b_{03}+b_{12})^2\right\}$.

The last two terms of above equations show that, the change of
energy density ($B^2 / \mu_g^g$) in the gravitational-gravitational
and electromagnetic-gravitational subfields will influence the
movement of particle \emph{m}. When the sum of last two terms in
above each equation is zero, the equations of
gravitational-gravitational subfield can be summarized as follows
\begin{subequations}
\begin{eqnarray}
&& \left\{(g_{01} + g_{23}) s_0^g + (g_{02} + g_{31}) s_3^g -
(g_{03} + g_{12})s_2^g \right\} + c (-\partial_{g0} s_1^g +
\partial_{g1} s_0^g + \partial_{g2}s_3^g - \partial_{g3}s_2^g) = 0
\label{a}\\
&& \left\{(g_{02} + g_{31}) s_0^g - (g_{01} + g_{23}) s_3^g +
(g_{03} + g_{12})s_1^g \right\} + c (-\partial_{g0} s_2^g  +
\partial_{g2}s_0^g + \partial_{g3}s_1^g - \partial_{g1} s_3^g) = 0
\label{b}\\
&&  \left\{(g_{03} + g_{12}) s_0^g + (g_{01} + g_{23}) s_2^g -
(g_{02} + g_{31})s_1^g \right\} + c (-\partial_{g0} s_3^g +
\partial_{g3}s_0^g + \partial_{g1} s_2^g - \partial_{g2}s_1^g ) = 0
\label{c}
\end{eqnarray}
\end{subequations}

From above equations we can learn that, when the sense of revolution
of particle \emph{m} is in the same sense of rotation of object
\emph{M}, the movement of particle \emph{m} is attracted by the
force which points to the orbit plane, hence the movement of
particle \emph{m} is stable. That is the near coplanarity of
planets. The particle \emph{m}, which moves around object \emph{M}
in spatial helix, revolves and waves up and down slowly around the
equator plane of object \emph{M} (the earth's movement relatives to
the equator plane of the sun; the movement of solar system and
pulsar relative to the galactic plane of the Milky Way Galaxy), at
the same time it rotates with swing \cite{sadler}. When the sense of
revolution of particle \emph{m} is in opposite sense of rotation of
object \emph{M}, the movement of particle \emph{m} is repelled by
the force which deviates from the orbit plane and hence the movement
of particle \emph{m} is unstable. It shows that, the sense of
revolution of particle \emph{m} affects directly its movement
stability. That is corevolving (or prograde) of planets. Therefore,
the solar system and galaxies are inclined to thin disk structure as
a whole \cite{kinney}.

To be extended, two galaxies rotating in the same sense of rotation
are in attraction and revolution around each other, even shorten
their interval and swallow up each other to become one galaxy. Two
galaxies rotating in opposite senses of rotation will repel and
increase their interval. Two galaxies rotating in arbitrary senses
of rotation will be affected by more complicated forces and moment
\cite{struck}.

\subsection{\label{sec:level2} Spatial helix movement}

On the orbit plane of particle \emph{m}, which angular momentum is
($M_0^g , M_1^g , M_2^g , M_3^g$), the gravitational-gravitational
subfield ($g_{01} + g_{23} , g_{02} + g_{31} , g_{03} + g_{12}$) is
generated by large-mass object \emph{M} which maintain its rotation
simultaneously in Fig.8. When the particle \emph{m}, which moves
around \emph{M} in spatial helix, stays in conservation of angular
momentum,
\begin{eqnarray}
\mathbb{W} = \left\{\emph{\textbf{i}}_1 (g_{01} + g_{23})+
\emph{\textbf{i}}_2 (g_{02} + g_{31}) + \emph{\textbf{i}}_3 (g_{03}
+ g_{12}) + \alpha \lozenge \right\} \circ (M_0^g +
\emph{\textbf{i}}_1 M_1^g + \emph{\textbf{i}}_2 M_2^g  +
\emph{\textbf{i}}_3 M_3^g)= 0
\end{eqnarray}
so its equations in $1 , \emph{\textbf{i}}_1 ,  \emph{\textbf{i}}_2
$ and $\emph{\textbf{i}}_3$ directions are respectively
\begin{subequations}
\begin{eqnarray}
&& c (\partial_{g0} M_0^g - \partial_{g1} M_1^g - \partial_{g2}
M_2^g - \partial_{g3} M_3^g) - \left\{(g_{01} + g_{23}) M_1^g +
(g_{02} + g_{31}) M_2^g + (g_{03} + g_{12}) M_3^g\right\} = 0
\label{a}\\
&& c (\partial_{g0} M_1^g + \partial_{g1} M_0^g +
\partial_{g2} M_3^g - \partial_{g3} M_2^g) + \left\{(g_{01}
+ g_{23}) M_0^g + (g_{02} + g_{31}) M_3^g - (g_{03} + g_{12}) M_2^g
\right\} = 0
\label{b}\\
&& c (\partial_{g0} M_2^g - \partial_{g1} M_3^g + \partial_{g2}
M_0^g + \partial_{g3} M_1^g ) + \left\{(g_{02} + g_{31}) M_0^g -
(g_{01} + g_{23}) M_3^g + (g_{03} + g_{12}) M_1^g \right\} = 0
\label{c}\\
&& c (\partial_{g0} M_3^g + \partial_{g1} M_2^g -
\partial_{g2} M_1^g + \partial_{g3} M_0^g ) + \left\{(g_{03} +
g_{12}) M_0^g + (g_{01} + g_{23}) M_2^g - (g_{02} +
g_{31}) M_1^g \right\} = 0 \label{D}
\end{eqnarray}
\end{subequations}

In the gravitational-gravitational subfield, above equations set
show that, the conservation of angular momentum will be affected by
the field strength, angular momentum and its variation rate etc.
When $M_1^g = M_2^g = 0$, we can obtain from the above equations set
\begin{subequations}
\begin{eqnarray}
&& \partial_{g3} M_3^g = \partial_{g0} M_0^g - \left\{(g_{03} +
g_{12}) M_3^g\right\}/c \label{a}\\
&& \partial_{g2} M_3^g = - \partial_{g1} M_0^g - \left\{(g_{01} +
g_{23}) M_0^g - (g_{02} + g_{31}) M_3^g \right\}/c \label{b}\\
&& \partial_{g1} M_3^g = \partial_{g2} M_0^g + \left\{(g_{02} +
g_{31}) M_0^g - (g_{01} + g_{23}) M_3^g \right\}/c \label{c}\\
&& \partial_{g0} M_3^g = - \partial_{g3} M_0^g - \left\{(g_{03} +
g_{12}) M_0^g \right\}/c \label{D}
\end{eqnarray}
\end{subequations}

When all terms in the right side of equal marks of the above
equations set are zero, the angular momentum $M_3^g$ of particle
\emph{m} along the sense of revolution is in conservation. And if
the right side of equal mark of the last equation is equal to zero
approximately, whilst the right side of equal marks of the first
three equations are not all zero, the angular momentum $M_3^g$ is
close to be in conservation and consequentially more complicated
situations will appear on the revolution orbit of particle \emph{m}.
For example, the asteroid will probably come near to the planet
gradually, and the moon may perhaps keep away from the earth in the
solar system \cite{prange, dzhunushaliev}.

\begin{figure}
\begin{minipage}[t]{0.5\linewidth}
\setcaptionwidth{3.3in}
\includegraphics[]{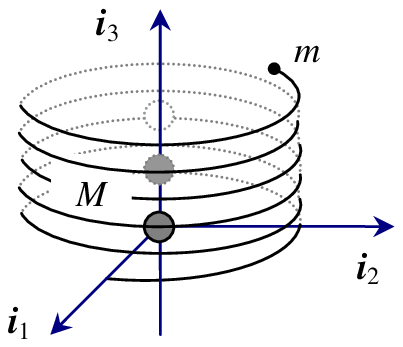}
\caption{Spatial helix movement. In Eq.(39), when the right side of
equal mark in last equation is equal to zero, and the first three
equations are not all zero, the angular momentum $M_3^g$ is close to
be in conservation and in consequence more complicated situations
will appear on the revolution orbit of particle \emph{m}.}
\label{fig:side:a}
\end{minipage}%
\begin{minipage}[t]{0.5\linewidth}
\setcaptionwidth{3.3in}
\includegraphics[]{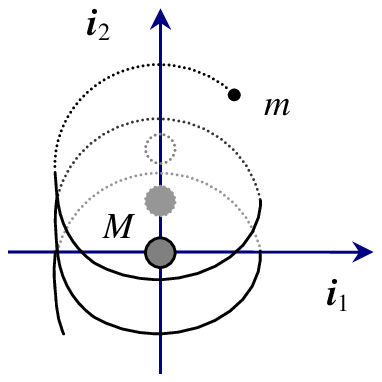}
\caption{Planar helix movement of celestial body. It shows that
there are the coupling influences between the 'dark matter field'
and 'modified gravitational field'. When the last term is zero in
Eq.(42), there will be no influence of the 'dark matter field' upon
the 'modified gravitational field'.} \label{fig:side:b}
\end{minipage}
\end{figure}

\subsection{\label{sec:level2} Planar helix movement of dark matter}

The field strength of gravitational-gravitational and
gravitational-electromagnetic subfields in Fig.9, which in equator
plane of large-mass object \emph{M}, are $(g_{01}, g_{02}, g_{12})$
and $(G_{01}, G_{02}, G_{12})$ respectively. The particle \emph{m}
moves around \emph{M} in planar helix, and its two types of
momentums in orbit plane are $(s_0^g , s_1^g , s_2^g , 0)$ and
$(s_0^e , s_1^e , s_2^e , 0)$ respectively. The orbit plane of
particle \emph{m} is superposed with equator plane of large-mass
object \emph{M}. When the particle \emph{m} is in force-balance
status,
\begin{eqnarray}
\mathbb{Z} = &&(\emph{\textbf{i}}_1 g_{01} + \emph{\textbf{i}}_2
g_{02} + \emph{\textbf{i}}_3 g_{12} + \emph{\textbf{I}}_1 G_{01} +
\emph{\textbf{I}}_2 G_{02} + \emph{\textbf{I}}_3 G_{12} + \alpha
\lozenge )^* \circ
\nonumber\\
&&\left\{\mu_g^g (s_0^g + \emph{\textbf{i}}_1 s_1^g +
\emph{\textbf{i}}_2 s_2^g) + \mu_g^e (\emph{\textbf{I}}_0 s_0^e
+\emph{\textbf{I}}_1 s_1^e + \emph{\textbf{I}}_2 s_2^e ) -
B^2/\alpha \right\} / \mu_g^g = 0
\end{eqnarray}
so its equations in $\emph{\textbf{i}}_1$ and $\emph{\textbf{i}}_2$
directions are respectively
\begin{subequations}
\begin{eqnarray}
&& \mu_g^g (g_{01} s_0^g - g_{12} s_2^g) + \alpha \mu_g^g
(-\partial_{g0} s_1^g + \partial_{g1} s_0^g - \partial_{g3} s_2^g)
\nonumber\\
&&+ \mu_g^e \left\{ \alpha (\partial_{e0} s_1^e -\partial_{e1} s_0^e
+ \partial_{e3} s_2^e) + (G_{12} s_2^e - G_{01} s_0^e) \right\} - (
g_{01} B^2 / \alpha + \partial_{g1} B^2) = 0 \label{a}
\\
&& \mu_g^g (g_{02} s_0^g + g_{12} s_1^g) + \alpha \mu_g^g
(-\partial_{g0} s_2^g +\partial_{g2} s_0^g + \partial_{g3} s_1^g)
\nonumber\\
&& + \mu_g^e \left\{ \alpha ( \partial_{e0} s_2^e -\partial_{e2}
s_0^e - \partial_{e3} s_1^e) - (G_{12} s_1^e + G_{02} s_0^e)
\right\} - ( g_{02} B^2 / \alpha +\partial_{g2} B^2) = 0   \label{b}
\end{eqnarray}
\end{subequations}
where, $B^2 = g_{01}^2 + g_{02}^2 + g_{12}^2 + G_{01}^2 + G_{02}^2
+G_{12}^2 $ .

When the last term of above each equation is zero approximately, we
can conclude
\begin{subequations}
\begin{eqnarray}
&& (g_{01} s_0^g - g_{12} s_2^g) + c (-\partial_{g0} s_1^g +
\partial_{g1} s_0^g - \partial_{g3} s_2^g)
\nonumber\\
&& + (\mu_g^e / \mu_g^g )\left\{ c (\partial_{e0} s_1^e -
\partial_{e1} s_0^e + \partial_{e3} s_2^e)
+ (G_{12} s_2^e - G_{01} s_0^e) \right\} = 0
\label{a} \\
&&(g_{02} s_0^g + g_{12} s_1^g) + c (-\partial_{g0} s_2^g +
\partial_{g2} s_0^g + \partial_{g3} s_1^g)
\nonumber\\
&& + (\mu_g^e / \mu_g^g )\left\{ c ( \partial_{e0} s_2^e -
\partial_{e2} s_0^e - \partial_{e3} s_1^e) - (G_{12} s_1^e + G_{02}
s_0^e) \right\} = 0
\label{b}
\end{eqnarray}
\end{subequations}

In above equations set, the above equations are the movement
equations of gravitational-gravitational subfield, including the
affecting from the gravitational-electromagnetic subfield. It
figures out that there are some coupling influences between the
'dark matter field' (gravitational-electromagnetic subfield) and
'modified gravitational field' (gravitational-gravitational
subfield). When the last term is zero in the above each equation,
there will be no influence of the 'dark matter field'
(gravitational-electromagnetic subfield) upon the 'modified
gravitational field' (gravitational-gravitational subfield).

If the preceding field strength and field source are extended to
four types of field strength and field source of
electromagnetic-gravitational field respectively, many extra and
also more complicated equations set of interplay between 'dark
matter field' and 'modified gravitational field', will be gained
from Eqs.(19) and (20).

\section{\label{sec:level1}CONCLUSIONS}

In the observed phenomena of the celestial body, which are
inconsistent with the results of current gravitational theory, some
are caused by the modified gravitational field
(gravitational-gravitational subfield) in the
electromagnetic-gravitational field, and some are caused by the dark
matter field (electromagnetic-electromagnetic and
gravitational-electromagnetic subfields).

The gravitational-gravitational subfield is the modified
gravitational field, which includes familiar Newtonian gravitational
field. The features of modified gravitational field predict that,
(1) the planetary orbits possess near coplanarity, near circularity
and corevolving, and the planets own rotation property, (2) the
centrifugal force of celestial body will change and lead to
fluctuation of revolution speed, when the field energy density or
the angular momentum of celestial body along the sense of revolution
varied, (3) the two galaxies rotating in the same sense of rotation
will attract and revolve around each other, hence shorten their
interval and swallow up each other to become one galaxy.

The electromagnetic-electromagnetic and
gravitational-electromagnetic subfields are both long range fields
and candidates of dark matter field. Their general charges are
candidates of dark matter. The combination of above general charges
and the general charges of gravitational-gravitational or
electromagnetic-gravitational subfields are candidates of dark
matter also. The field strength of electromagnetic-electromagnetic
and gravitational-electromagnetic subfields may be equal and very
weak, and a little less than that of gravitational-gravitational
subfield. The features of dark matter predict that, (1) the dark
matter field possesses two sorts of general charges makes the dark
matter particles very diversiform, (2) the field strength of two
types of dark matter fields are both weaker than that of the
Newtonian gravitational filed, (3) the distribution of dark matter
will affect the stabilizing velocity of celestial body. If the field
sources of the electromagnetic-electromagnetic and
gravitational-electromagnetic subfields distribute properly, the
obtained gravity of celestial body will increase, and the
stabilizing velocity of celestial body will also increase
correspondingly.

\begin{acknowledgments}
This project was supported partly by the National Natural Science
Foundation of China under grant number 60677039, Science \&
Technology Department of Fujian Province of China under grant number
2005HZ1020 and 2006H0092, and Xiamen Science \& Technology Bureau of
China under grant number 3502Z20055011.
\end{acknowledgments}

\end{document}